\documentclass{elsart}
\usepackage{graphicx, amssymb}
\journal{Physica A}

\begin{document}

\begin{frontmatter}



\title{Exact results of a mixed spin-1/2 and spin-$S$ Ising model on a bathroom tile (4-8) lattice: \\
effect of uniaxial single-ion anisotropy\thanksref{VEGA}}
\thanks[VEGA]{This work was financially supported under the grants VEGA 1/2009/05 and APVT 20-005204.}

\author{Jozef Stre\v{c}ka}
\ead{jozkos@pobox.sk}
\address{Department of Theoretical Physics and Astrophysics, 
Faculty of Science, \\ P. J. \v{S}af\'{a}rik University, Park Angelinum 9,
040 01 Ko\v{s}ice, Slovak Republic}

\begin{abstract}
Effect of uniaxial single-ion anisotropy upon magnetic properties of a mixed spin-1/2 and spin-$S$ 
($S \geq 1$) Ising model on a bathroom tile (4-8) lattice is examined within the framework of an exact 
star-triangle mapping transformation. The particular attention is focused on the phase diagrams established for several values of the quantum spin number $S$. It is shown that the mixed-spin bathroom tile lattice exhibits very similar phase boundaries as the mixed-spin honeycomb lattice whose critical points are merely slightly enhanced with respect to the former ones. The influence of uniaxial single-ion anisotropy upon the total magnetization vs. temperature dependence is particularly investigated as well. 
\end{abstract}


\end{frontmatter}

\section{Introduction}

Over the last six decades, much effort has been devoted to determine a criticality and other statistical properties of various lattice-statistical models, which would enable a deeper understanding of order-disorder phenomena in solids. The planar Ising model takes a prominent place in the equilibrium statistical mechanics from this point of view as it represents a rare example of exactly solvable lattice-statistical model since Onsager's pioneering work \cite{onsager}. It is worthy to notice, however, that exactly soluble planar Ising models \cite{baxter} are of immense interest because of providing a convincing evidence for many controversial results predicted in the phase transition theory and moreover, they also provide a good testing ground for  approximative theories. Nevertheless, it should be mentioned that a precise treatment of planar Ising 
models is usually connected with the usage of sophisticated mathematical methods, which consequently lead to considerable difficulties when applying them to more complicated but simultaneously more realistic models.   

The Ising systems consisting of mixed spins of different magnitudes, so-called {\it mixed-spin Ising models}, being among the most interesting extensions of the standard spin-1/2 Ising model which are currently of  great research interest. These models have recently enjoyed a considerable attention chiefly due to much richer critical behavior they display compared with their single-spin counterparts. Magnetic properties of the mixed-spin Ising models can essentially be modified, for instance, by a presence of single-ion anisotropy acting on the $S \geq 1$ spins. Indeed, this effect may potentially cause a tricritical phenomenon -- a change of the character of phase transition from a second-order to a first-order one. Another aspect which started to attract an appreciable interest towards the mixed-spin Ising models can be related to 
a theoretical modeling of magnetic structures suitable for describing a ferrimagnetism of certain class of insulating magnetic materials. In this respect, the ferrimagnetic mixed-spin Ising models are very 
interesting also from the experimental point of view, first of all in connection with many possible technological applications of ferrimagnets to practice (e.g. thermomagnetic recording).

Despite the intensive studies, there are only few examples of exactly solvable mixed-spin Ising models yet. 
Using the generalized form of decoration-iteration and star-triangle transformations, the mixed spin-1/2 
and spin-$S$ ($S \geq 1$) Ising models on the honeycomb, diced and decorated honeycomb lattices were 
exactly been treated by Fisher long ago \cite{MT}. Notice that these mapping transformations were later on further generalized in order to account also for the single-ion anisotropy effect. It is worth mentioning 
that in this way generalized mapping transformations were recently employed to obtain exact results of the mixed-spin Ising models on the honeycomb lattice \cite{goncalves} as well as on some decorated planar 
lattices \cite{jascur}. To the best of our knowledge, these are the only mixed-spin planar Ising models with generally known exact solutions except several mixed-spin Ising models on the Bethe (Cayley tree) 
lattices studied within a discrete non-linear map \cite{bethe1} and an approach based on exact recursion equations \cite{bethe2}. Among the remarkable models for which a precise solution is restricted to 
a certain subspace of interaction parameters only, one should further mention the mixed-spin Ising model on the Union-Jack (centered square) lattice treated within the mapping onto a symmetric eight-vertex model \cite{union}.  

The purpose of this work is to provide an exact formulation of the mixed spin-1/2 and spin-$S$ ($S \geq 1$) Ising model on the bathroom tile (4-8) lattice and to establish accurate phase diagrams of this system 
for several values of the quantum spin number $S$. Exact results for the system under consideration are obtained by applying the generalized star-triangle mapping transformation constituting an exact 
correspondence with an effective spin-1/2 Ising model on the Shastry-Sutherland (orthogonal-dimer) lattice (further abbreviated as SSL), which has recently been solved by the present author elsewhere \cite{ssl}. The obtained phase diagrams will also be compared with those of another exactly soluble three-coordinated planar Ising model, namely, on the mixed-spin honeycomb lattice \cite{goncalves}. Next, the influence of uniaxial single-ion anisotropy upon the temperature dependence of the total magnetization will be clarified. 

The outline of this paper is as follows. In Section 2, a detailed formulation of the considered model 
is presented and subsequently, exact expressions for the magnetization and critical temperatures are 
derived within the exact mapping onto the effective spin-1/2 Ising model on the anisotropic SSL. 
The most interesting numerical results obtained for several values of the quantum spin number $S$ are 
presented and particularly discussed in Section 3. Finally, some concluding remarks are given in Section 4.

\section{Model and its exact solution}

Let us begin by considering a mixed spin-1/2 and spin-$S$ ($S \geq 1$) Ising model on the bathroom tile (4-8) lattice schematically illustrated on the left-hand side of Fig. 1. The mixed-spin 4-8 lattice consists of two interpenetrating inequivalent sub-lattices $A$ and $B$, which are formed by the spin-1/2 (empty circles) and   spin-$S$ (solid circles) atoms, respectively. The Ising Hamiltonian defined upon the underlying 4-8 lattice reads:
\begin{eqnarray}
{\mathcal H}_{4-8} = - J \sum_{(i,j) \subset \mathcal L}^{3N} S_{i}^{A} S_{j}^{B} 
                     - D \sum_{i=1}^{N} (S_{i}^{B})^2,     
\label{H48}
\end{eqnarray}
where $S_i^A = \pm 1/2$ and $S_i^B = -S, -S+1, ..., S$ are the Ising spins located at $i$th lattice point, 
their superscript denotes the sub-lattice to which they belong, $J$ being the exchange interaction between nearest-neighboring $A-B$ spin pairs and the parameter $D$ measures a strength of the uniaxial single-ion 
anisotropy acting on the spins of sub-lattice $B$. Furthermore, $N$ denotes the total number of sites at each sub-lattice, the first summation is carried out over all nearest-neighboring $A-B$ spin pairs and the 
second summation runs over all sites of sub-lattice $B$.

It is of principal importance that the 4-8 lattice belongs to 'loose-packed' lattices, i.e. it can be 
divided into two interpenetrating sub-lattices $A$ and $B$ in such a way that all nearest neighbors of $A$-sites belong to the sub-lattice $B$ and vice versa. Owing to this fact, a summation over the spins 
of one sub-lattice (say sub-lattice $B$) can be performed before summing over the spins of another sub-lattice (sub-lattice $A$) when developing the partition function of the system under investigation. In the consequence of that, the partition function of the mixed-spin 4-8 lattice can be rewritten in this useful form:
\begin{eqnarray}
{\mathcal Z}_{4-8} = \sum_{\{S^{A} \}} \prod_{i=1}^{N} \sum_{S^{B}_{i} = -S}^{+S}
                     \exp[\beta D (S^B_i)^2] \exp[\beta J S^B_i (S^A_j + S^A_k + S^A_l)],     
\label{Z48}
\end{eqnarray}
where $\beta = 1/(k_{\mathrm B} T)$, $k_{\mathrm B}$ being Boltzmann's constant, $T$ stands for the absolute temperature, the product is taken over all $B$-sites and the symbol $\sum_{\{S^{A} \}}$ means a summation 
over all possible spin configurations of the sub-lattice $A$. It is quite obvious that this form justifies 
a performance of the familiar star-triangle mapping transformation \cite{MT}:
\begin{eqnarray}
\sum_{S^{B}_{i} = -S}^{S} && \exp[\beta D (S^B_i)^2] \exp[\beta J S^B_i (S^A_j + S^A_k + S^A_l)] \nonumber \\
&&  = C \exp[\beta J_{\rm inter} (S^A_j S^A_k + S_k^A S^A_l + S^A_l S^A_j)],     
\label{ST}
\end{eqnarray}
which establishes an exact correspondence between the mixed-spin Ising model on the 4-8 lattice and 
its equivalent spin-1/2 Ising model on the SSL when applying (\ref{ST}) to all $B$-type sites (Fig. 1). Notice 
that the unknown mapping parameters $C$ and $J_{\rm inter}$ are unambiguously given directly by the 
transformation (\ref{ST}), which must hold for all available configurations of $S_j^A$, $S^A_k$, $S^A_l$ spins. According to this, one readily finds:
\begin{eqnarray}
C = V_1^{1/4} V_2^{3/4}, \qquad \quad \beta J_{\rm inter} = \ln(V_1/V_2), 
\label{MP}
\end{eqnarray}
with the expressions $V_1$ and $V_2$ defined as:
\begin{eqnarray}
V_1 &=&  \sum_{n=-S}^{S} \exp(\beta D n^2) \cosh(3 \beta J n/2),  \label{MP1}   \\
V_2 &=&  \sum_{n=-S}^{S} \exp(\beta D n^2) \cosh(\beta J n/2).    \label{MP2}
\end{eqnarray}

At this stage, let us substitute the mapping (\ref{ST}) into the formula (\ref{Z48}) gained for 
the partition function of mixed-spin 4-8 lattice. After straightforward rearrangement, 
one easily obtains an exact relationship between the partition function ${\mathcal Z}_{4-8}$ of the 
mixed-spin 4-8 lattice and its corresponding partition function ${\mathcal Z}_{SSL}$ of the 
spin-1/2 Ising model on the anisotropic SSL:
\begin{eqnarray}
{\mathcal Z}_{4-8} (\beta, J, D) = C^N {\mathcal Z}_{SSL} (\beta, J_{\rm intra}, J_{\rm inter}),
\label{PF}
\end{eqnarray}
definitely given by the effective Hamiltonian:
\begin{eqnarray}
{\mathcal H}_{SSL} = - J_{\rm intra} \sum_{(i,j) \subset \mathcal D}^{N/2} S_{i}^{A} S_{j}^{A} 
                     - J_{\rm inter} \sum_{(i,j) \subset \mathcal C}^{2N} S_{i}^{A} S_{j}^{A}.     
\label{HSSL}
\end{eqnarray}
Above, the first (second) summation accounts nearest-neighbor intra-dimer (inter-dimer) interactions 
of the SSL that are displayed on the right-hand side of Fig. 1 as solid (dashed) lines and simultaneously,  
the condition $J_{\rm intra} = 2 J_{\rm inter}$ must be satisfied. With regard to this, the isotropic mixed-spin 4-8 lattice is mapped onto the anisotropic spin-1/2 Ising model on the SSL recently solved by 
the present author elsewhere \cite{ssl}. The intra-dimer coupling of the SSL is twice as large as the inter-dimer one because the former interaction is being counted twice when performing the star-triangle 
transformation (\ref{ST}) at both opposite $B$-sites of the same square plaquette of the original 4-8 lattice, 
while the latter one is being counted just once. 

The mapping relation (\ref{PF}) between the partition functions ${\mathcal Z}_{4-8}$ and ${\mathcal Z}_{SSL}$ represents a central result of our calculation, in fact, it can be utilized for establishing similar mapping relations also for other important quantities such as Gibbs free energy, internal energy, magnetization, correlation functions, specific heat, etc. When combining the formula (\ref{PF}) with commonly used 
mapping theorems \cite{barry}, one easily proves a validity of following relations:
\begin{eqnarray}
m_A &\equiv& \langle S_j^A \rangle_{4-8} = \langle S_j^A \rangle_{SSL} \equiv m_{SSL};  
\label{mA} \\
t_A &\equiv& \langle S_j^A S_k^A S_l^A \rangle_{4-8} = \langle S_j^A S_k^A S_l^A \rangle_{SSL} \equiv t_{SSL}; 
\label{tA} \\
m_B  &\equiv& \langle S_i^B \rangle_{4-8} = 3 m_A [F(3) + F(1)]/2 + 2 t_A [F(3) - 3F(1)];                   
\label{mB}
\end{eqnarray}
where the symbols $\langle ... \rangle_{4-8}$ and $\langle ... \rangle_{SSL}$ denote standard canonical average performed on the Ising lattices defined by means of the Hamiltonian (\ref{H48}) and (\ref{HSSL}), respectively, and the function $F(x)$ labels expression:
\begin{eqnarray}
F(x) =  \frac{\displaystyle \sum_{ n=-S}^{S} \exp(\beta D n^2) n \sinh(\beta J n x/2)}
             {\displaystyle \sum_{n=-S}^{S} \exp(\beta D n^2)   \cosh(\beta J n x/2)}. 
\label{F}
\end{eqnarray}
It should be pointed out that the set of Eqs. (\ref{mA})-(\ref{mB}) enables straightforward calculation of both sub-lattice magnetization $m_A$ and $m_B$ of the mixed-spin Ising model on the 4-8 lattice. According to 
Eq. (\ref{mA}), the sub-lattice magnetization $m_A$ directly equals to the corresponding magnetization  $m_{SSL}$ of the SSL being unambiguously given by the mapping relation (\ref{MP}) and the constraint 
$J_{\rm intra} = 2 J_{\rm inter}$. On the other hand, the triplet order parameter $t_{SSL}$ is required 
for a computation of the sub-lattice magnetization $m_B$ in addition to $m_{SSL}$. Although a precise 
evaluation of the triplet correlation $t_{SSL}$ between three spins located within the triangular 
plaquettes of the SSL has not been presented in Ref. \cite{ssl}, it can be accomplished following 
the same procedure as presented there for $m_{SSL}$. Subsequently, one readily proves that the triplet 
correlation $t_{SSL}$ can be expressed in terms of the magnetization $m_{SSL}$ only. In view of these 
facts, an exact solution for both sub-lattice magnetization of the mixed-spin Ising model on the 
4-8 lattice is thus formally completed.  

Finally, an analytical condition determining exact critical points of the mixed-spin 4-8 lattice will be derived. It can easily be understood that both sub-lattice magnetization $m_A$ and $m_B$ vanish when
the magnetization $m_{SSL}$ of the corresponding SSL tends to zero. In order to provide an accurate 
condition for the criticality of mixed-spin 4-8 lattice it is therefore sufficient to ascertain 
an exact critical point of the anisotropic SSL fulfilling the claim $J_{\rm intra} = 2 J_{\rm inter}$. 
After straightforward but a little bit tedious modify based on the procedure presented in Ref. \cite{ssl}, 
one finds an exact critical point of the anisotropic SSL with this special ratio of inter- and intra-dimer coupling constants being:
\begin{eqnarray}
\beta_c J_{\rm intra} = \frac{J_{\rm intra}}{k_{\rm B} T_c} =  
2 \ln \biggl( \frac{8 + 5 \sqrt{2} + \sqrt{10 + 8 \sqrt{2}} + 2 \sqrt{5 + 4 \sqrt{2}}}
                  {2 + \sqrt{2} + \sqrt{10 + 8 \sqrt{2}}} \biggr). 
\label{tcssl}
\end{eqnarray}
Regarding this, a direct substitution of the critical point (\ref{tcssl}) into the relevant mapping relation (\ref{MP}) immediately yields the critical condition determining phase boundaries of the mixed-spin Ising model on the 4-8 lattice given by:
\begin{eqnarray}
&& \frac{\displaystyle \sum_{n=-S}^{S} \exp(\beta_c D n^2) \cosh(3 \beta_c J n/2)}
        {\displaystyle \sum_{n=-S}^{S} \exp(\beta_c D n^2) \cosh(\beta_c J n/2)} =  \nonumber \\
&& \frac{8 + 5 \sqrt{2} + \sqrt{10 + 8 \sqrt{2}} + 2 \sqrt{5 + 4 \sqrt{2}}}
        {2 + \sqrt{2} + \sqrt{10 + 8 \sqrt{2}}}. 
\label{tc48}
\end{eqnarray}

\section{Results and Discussion}

Let us step forward to a discussion of the most interesting numerical results. Although all results 
presented in the preceding section hold for the ferromagnetic ($J > 0$) as well as ferrimagnetic ($J < 0$) version of the model under investigation, in what follows we shall restrict ourselves to an analysis 
of the ferrimagnetic model only. It appears worthwhile to mention, nevertheless, that phase 
diagrams displayed below for the ferrimagnetic model are maintained without any changes also for 
the ferromagnetic model as a result of an invariance of the critical condition (\ref{tc48}) with 
respect to the transformation $J \to -J$.

The uniaxial single-ion anisotropy basically modifies the magnetic behavior of the spin-$S$ atoms 
that constitute the sub-lattice $B$ of mixed-spin 4-8 lattice. To illustrate the case, our attention 
was firstly focused on the ground-state behavior explored for arbitrary quantum spin number $S$. When 
the strength of the easy-plane single-ion anisotropy ($D < 0$) reinforces, the system exhibits successive first-order phase transitions associated with lowering the spin moment $S^B$ from its highest possible value. The cascade of the uniaxial single-ion anisotropies at which these first-order transitions 
occur is given by:
\begin{eqnarray}
\frac{D_c}{|J|} = \frac{-3}{4n-2},
\label{gs}
\end{eqnarray} 
where $n = 3/2, 5/2, ..., S$ for half-integer spin-$S$ atoms, while $n = 1, 2, ..., S$ for integer 
spin-$S$ atoms. The most fundamental difference between the mixed-spin systems with half-integer 
spins placed on sub-lattice $B$ and respectively, their integer spins counterparts, consists in an 
appearance of the disordered phase in the former systems due to a presence of the 'non-magnetic' state 
$S^{B} = 0$ below $D_c/|J| = -1.5$. The latter systems, on the other hand, remain long-range ordered 
even below their last critical value $D_c/|J| =-0.75$, where their lowest but 'magnetic' spin 
state $S^{B} = 1/2$ becomes stable.  

Now, we turn to a detailed investigation of finite-temperature phase diagrams as depicted in Fig. 2A 
(Fig. 2B) demonstrating the variations of critical temperatures with the uniaxial single-ion anisotropy 
for several values of half-integer (integer) spins constituting the sub-lattice $B$. In these figures, 
the critical temperatures of the mixed-spin Ising model on the 4-8 lattice (solid lines) are compared 
with critical lines of the mixed-spin Ising model on the honeycomb lattice (dotted lines) that represents another exactly soluble three-coordinated planar Ising lattice \cite{goncalves}. It can be clearly seen from Fig. 2 that both the mixed-spin Ising lattices exhibit very similar dependence on the single-ion anisotropy strength as the critical temperatures of the former lattice are merely slightly reduced with respect to those of the latter one. This result provides a further confirmation of the conjecture, originally supposed from a comparison of critical points of the spin-1/2 Ising model on the square and kagom\'e lattices \cite{sqka}, that the critical points of Ising lattices with the same coordination number but different lattice topology being slightly lower for the lattices composed of more regular polygons. Finally, it is worthy to notice 
that the critical temperatures shown in Fig. 2A correctly reproduce an exact critical point of the isotropic spin-1/2 Ising model on the 4-8 lattice \cite{48} by considering the limiting case $D/|J| \to -\infty$.   
 
For easy reference, let us introduce the classification scheme diagrammatically shown in Fig. 3 before proceeding to a discussion of the most interesting results obtained for temperature dependences of the resultant magnetization. It seems unavoidable that the N\'eel's nomenclature\footnotemark referring to 
ferrimagnetic $m(T)$ curves should be furnished by several less frequent but notable dependences. 
In addition to the familiar Q-, R-, P- and N-type curves, which are commonly found in the standard textbooks 
on magnetism \cite{chika}, the L-, M-, S- and W-type curves are added here to this basic classification scheme. \footnotetext[1]{L. N\'eel \cite{neel} originally envisaged 6 distinct types of $m(T)$ curves, 
but some of them having a non-zero initial slope in contradiction with the $m(T)$ curves commonly found
in the ferrimagnetic Ising systems. Therefore, the nomenclature used further is based on a set of the 
standard Q-, R-, P- and N-type curves \cite{chika} filled out by L-, M-,  S- and W-type dependences.}
The conventional Q- and R-type dependences show a monotonic decrease of the magnetization with increasing 
the temperature, the former one displaying a steep decrease of the magnetization just in the vicinity of critical temperature, while the latter one already exhibits a relatively rapid magnetization decline 
within the range of intermediate temperatures before going abruptly to zero at the critical point. 
Another familiar P-type dependence shows the temperature-induced maximum as the temperature raises, 
whereas the N-type curve is being characterized by one compensation point at which resultant magnetization disappears due to the complete cancellation of the sub-lattice magnetization. The L-type curve \cite{barbara} 
is very analogous to the P-type dependence, however, the resultant magnetization starts from zero in this particular case. Similarly, the M-type dependence also begins from zero but it shows two separate maxima 
before reaching the critical temperature. The stair-like S-shaped dependence initially exhibits a steep decrease of the magnetization that almost completely diminishes in the range of intermediate temperatures 
and finally, a second steep decrease follows near the critical temperature. Last, the striking W-type 
curve peculiarly exhibits two compensation points before the critical one.

The temperature dependences of resultant magnetization normalized per one site of the mixed-spin 4-8 
lattice ($m = |m_A + m_B|/2$) are depicted in Fig. 4. The total magnetization is plotted against the 
temperature for several values of the quantum spin number $S$ and various single-ion anisotropies. 
At first, let us look more closely on the particular cases when the sub-lattice $B$ consists of 
the integer spin-$S$ atoms. As far as the mixed spin-1/2 and spin-1 system is considered, the standard 
Q-type curve observable for $D \geq 0$ gradually changes into the R-shaped dependence by reversing 
the sign of single-ion anisotropy term. The shape of the R-type dependence can be attributed to more rapid 
thermal variation of the sub-lattice magnetization $m_B$, which is thermally more easily disturbed than 
$m_A$ under the condition $D < 0$. The situation becomes even a little bit more involved when assuming the mixed spin-1/2 and spin-2 system. Before entering the disordered phase at $D_c/|J| = -1.5$, this system exhibits in the ground state another first-order phase transition between two spontaneously ordered 
phases having the spin-2 atoms in the states $S^B = 2$ and $1$, respectively. 
Consequently, slightly below the critical value $D_c/|J| = -0.5$ that determines the coexistence of 
both the ordered phases one also encounters the P-shaped dependence with the temperature-induced 
maximum arising from $S^B = 1 \to 2$ thermal excitations (see the curve labeled $D/|J| = -0.6$). 
By contrast, the opposite $S^B = 2 \to 1$ thermal excitations are responsible for 
the occurrence of the S-shaped dependence when selecting the single-ion anisotropy strength 
slightly above this critical value (see for instance $D/|J| = -0.45$).

Before concluding, the most remarkable features of the mixed-spin 4-8 lattice with the half-integer 
spin-$S$ atoms on the sub-lattice $B$ will be summarized. It should be stressed that the systems 
also exhibit the Q-, R-, P- as well as S-shaped curves when varying the single-ion anisotropy strength 
(see the right panel of Fig. 4). The curvature of these dependences is apparently caused by the 
same reasons as discussed earlier for their integer-spin counterparts and hence, the discussion concerning 
with their origin  being here omitted for brevity. It is worthy to recall, however, that the mixed-spin 
systems with half-integer spin-$S$ atoms possess the antiferromagnetically ordered ground state below
their last critical value $D_c/|J| = -0.75$. On account of the preferable thermal excitations 
$S^B = 1/2 \to 3/2$ and $S^B = 1/2 \to 5/2$, respectively, the L-type curve develops for single-ion anisotropies close enough to this critical value (e.g. $D/|J| = -0.8$ and $-1.0$). Notice that 
the L-shaped dependence should not be confused neither with the magnetization of systems exhibiting 
reentrant transitions (in our case, the total magnetization rises much more steadily near the zero temperature), nor with the $m(T)$ curves of ordinary antiferrromagnets placed in a non-zero external 
magnetic field. Finally, it is noteworthy that the mixed-spin systems having the half-integer spin-$S$ 
atoms on the sub-lattice $B$ indeed behave, due to the almost complete cancellation of antiferromagnetically arranged spins, as ordinary antiferromagnets when $D/|J| \lesssim -1.5$. In this respect,
the critical frontiers displayed in Fig. 2A should be better considered in this range of parameters 
as the critical boundaries determining disappearance of the spontaneous 'antiferromagnetic' order.

\section{Concluding Remarks}

In the present article, the star-triangle mapping is being used for computing the exact phase diagrams
and magnetization of the mixed spin-1/2 and spin-$S$ ($S \geq 1$) Ising model on the bathroom tile (4-8) lattice. The particular attention was focused on the influence of uniaxial single-ion anisotropy on the criticality and the shape of the magnetization vs. temperature dependence. Depending on the quantum spin number $S$ and the single-ion anisotropy strength, the temperature variation of resultant magnetization 
has been found to be either of Q-, R-, P-, L- or S-type. The N- and W-type  curves with one 
or two compensation temperatures, respectively, cannot appear in the system under investigation on behalf 
of the uniaxial single-ion anisotropy effect. 

It is worthwhile to mention that the relative ease of the calculation procedure presented here enables 
further remarkable extensions. Actually, the approach based on the star-triangle mapping transformation 
can appreciably be developed to account also for: i) different nearest-neighbor interactions assigned to 
the bonds within the square plaquettes of the original 4-8 lattice and respectively, to the links between them; ii) second-neighbor interactions between the $A$-sites; iii) four-spin interaction between each 
$B$-site and its three nearest-neighboring $A$-sites; iv) the biaxial single-ion anisotropy acting 
on the $B$-sites. Notice that the last modification necessitates following the extended version of mapping procedure worked out previously for the mixed spin-1/2 and spin-$S$ ($S \geq 1$) Ising model on the 
honeycomb lattice \cite{biaxial}. In the spirit of results reported on formerly (see e.g. Refs. \cite{jascur} and \cite{biaxial}), it is quite reasonable to suppose that the extensions ii)-iv) may significantly 
modify the magnetic behavior of studied systems and leastwise the N- and W-type curves should be 
expected to emerge in these modified versions of mixed-spin Ising models. 



\newpage
\begin{large}
\textbf{Figure captions}
\end{large}

\begin{itemize}

\item [Fig. 1]
Diagrammatic representation of the mapping scheme. The isotropic mixed-spin bathroom tile (4-8) lattice consisting of two inequivalent sub-lattices $A$ and $B$ constituted by the spin-1/2 (empty circles) and spin-$S$ (solid circles) atoms, respectively, is mapped on the anisotropic spin-1/2 Ising 
model on the Shastry-Sutherland (orthogonal-dimer) lattice with the intra- and inter-dimer interactions 
schematically illustrated as solid and dashed lines. 

\item [Fig. 2]
Phase diagrams of the mixed-spin Ising model on the bathroom tile (solid lines) and honeycomb (dotted lines) lattices. Figure on the right (left) illustrates critical frontiers of the mixed-spin systems composed 
of the integer (half-integer) spin-$S$ atoms. The empty circles demarcate special critical points 
of the mixed-spin 4-8 lattice that correspond to the coexistence of two different magnetically ordered phases.

\item [Fig. 3]
Diagrammatic representation of various types of the ferrimagnetic magnetization vs. temperature 
dependences. Acronyms for particular cases are given by the circled letters.

\item [Fig. 4]
The temperature variation of resultant magnetization of the mixed-spin 4-8 lattice exposed for several 
values of quantum spin number $S$ and diverse uniaxial single-ion anisotropy strengths.  

\end{itemize} 


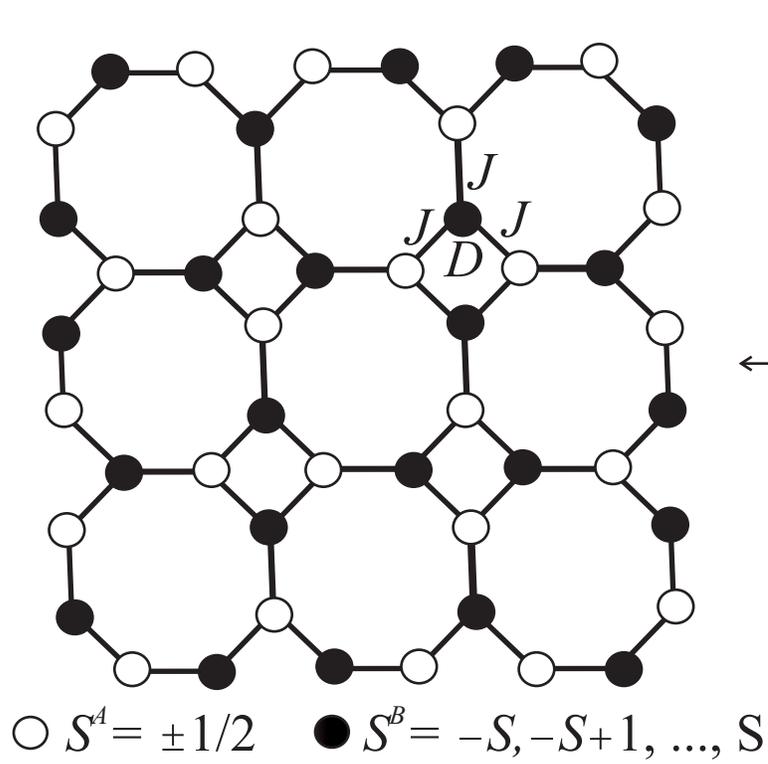
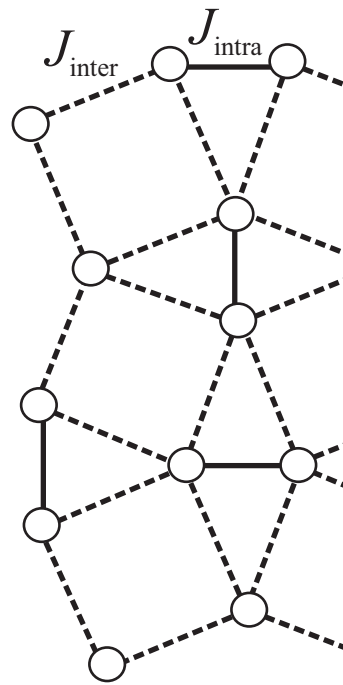
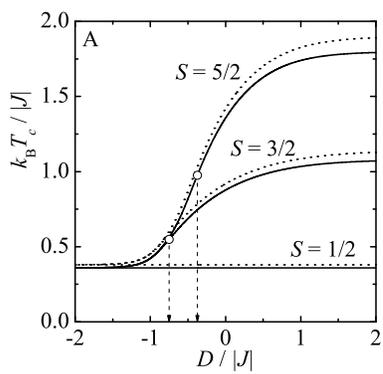
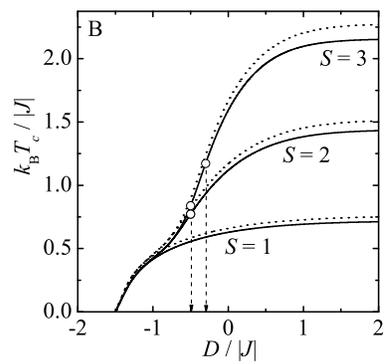
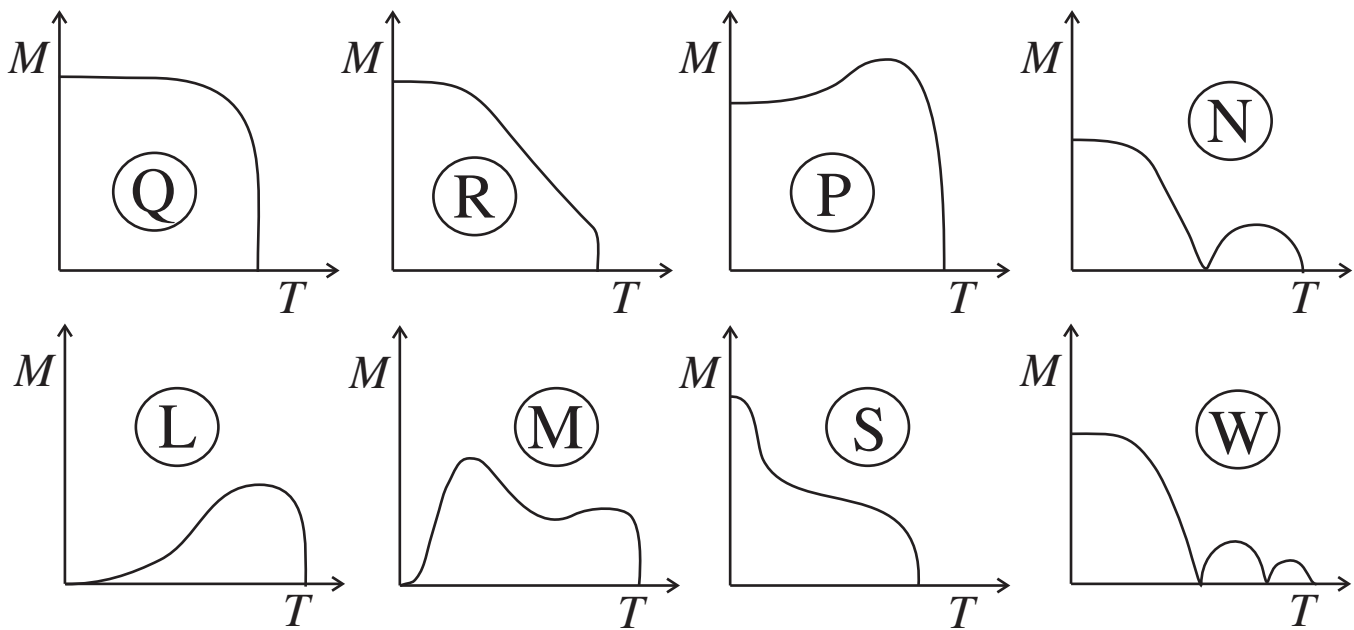
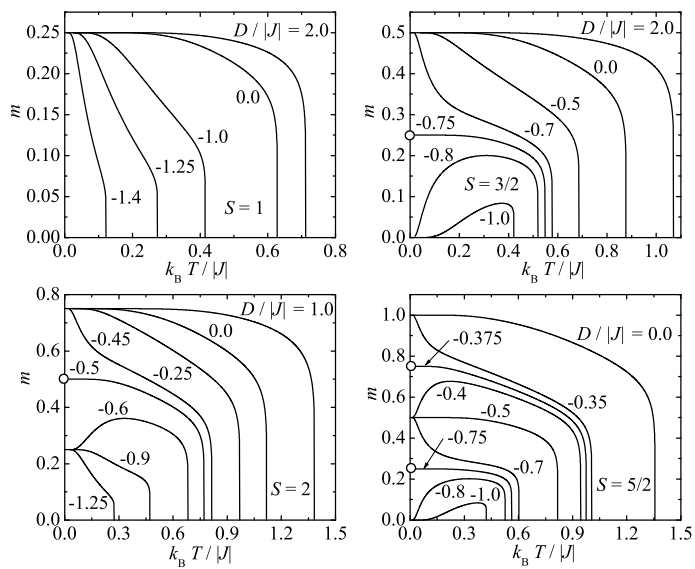
\end{document}